\documentstyle[12pt,epsfig]{article}
\def\be{\begin{equation}}
\def\ee{\end{equation}}
\def\bc{\begin{center}}
\def\ec{\end{center}}
\def\bea{\begin{eqnarray}}
\def\eea{\end{eqnarray}}

\def\nn{\nonumber}

\def\auxz{{F^Z}}

\def\cl{{\cal L}}

\def\ev{{\rm \; eV}}

\def\gev{{\rm \; GeV}}

\def\grav{\tilde{G}}
\def\gt{\tilde{G}}

\def\mgg{M_{\gamma \gamma}}
\def\mgz{M_{\gamma Z}}

\def\mzz{M_{ZZ}}

\def\ov{\overline}
\def\pb{{\; \rm pb}}

\def\simlt{\stackrel{<}{{}_\sim}}

\def\tev{{\rm \; TeV}}

\catcode`@=11
\def\marginnote#1{}
\newcount\hour
\newcount\minute
\newtoks\amorpm
\hour=\time\divide\hour by60
\minute=\time{\multiply\hour by60 \global\advance\minute by-\hour}
\edef\standardtime{{\ifnum\hour<12 \global\amorpm={am}%
        \else\global\amorpm={pm}\advance\hour by-12 \fi
        \ifnum\hour=0 \hour=12 \fi
        \number\hour:\ifnum\minute<10 0\fi\number\minute\the\amorpm}}
\edef\militarytime{\number\hour:\ifnum\minute<10 0\fi\number\minute}
\def\draftlabel#1{{\@bsphack\if@filesw {\let\thepage\relax
   \xdef\@gtempa{\write\@auxout{\string
      \newlabel{#1}{{\@currentlabel}{\thepage}}}}}\@gtempa
   \if@nobreak \ifvmode\nobreak\fi\fi\fi\@esphack}
        \gdef\@eqnlabel{#1}}
\def\@eqnlabel{}
\def\@vacuum{}
\def\draftmarginnote#1{\marginpar{\raggedright\scriptsize\tt#1}}
\def\draft{\oddsidemargin 0.0truein
        \def\@oddfoot{\sl preliminary draft \hfil
        \rm\thepage\hfil\sl\today\quad\militarytime}
        \let\@evenfoot\@oddfoot \overfullrule 3pt
        \let\label=\draftlabel
        \let\marginnote=\draftmarginnote
   \def\@eqnnum{(\theequation)\rlap{\kern\marginparsep\tt\@eqnlabel}%
\global\let\@eqnlabel\@vacuum}  }
\catcode`@=12
\begin{document}
\begin{titlepage}
\vspace*{-1cm}
\phantom{hep-ph/0001025} 
\hfill{DFPD-99/TH/55}
\vskip 2.0cm
\begin{center}
{\Large\bf Signatures of massive sgoldstinos at $e^+ 
e^-$ colliders}
\end{center}
\vskip 1.5  cm
\begin{center}
{\large Elena 
Perazzi}\footnote{e-mail address: perazzi@pd.infn.it}
\\
\vskip .1cm
Dipartimento di Fisica, Universit\`a di Padova, I-35131 Padua, Italy
\\
\vskip .2cm
{\large Giovanni Ridolfi}\footnote{e-mail address: ridolfi@ge.infn.it}
\\
\vskip .1cm
INFN, Sezione di Genova, I-16146 Genova, Italy
\\
\vskip .2cm
and
\\
\vskip .2cm
{\large Fabio
Zwirner}\footnote{e-mail address: zwirner@pd.infn.it}
\\
\vskip .1cm
INFN, Sezione di Padova, I-35131 Padua, Italy
\end{center}
\vskip 2.5cm
\begin{abstract}
\noindent
In supersymmetric extensions of the Standard Model with a very
light gravitino, the effective theory at the weak scale should 
contain not only the goldstino $\gt$, but also its supersymmetric 
partners, the sgoldstinos. In the simplest case, the goldstino 
is a gauge-singlet and its superpartners are two neutral spin--0 
particles, $S$ and $P$. We study possible signals of massive 
sgoldstinos at $e^+ e^-$ colliders, focusing on those that are 
most relevant at LEP energies. We show that the LEP constraints 
on $e^+ e^- \rightarrow \gamma S \, (\gamma P)$, $Z S \, (Z P)$ 
or $e^+ e^- S \, (e^+ e^- P)$, followed by $S (P)$ decaying into 
two gluon jets, can lead to stringent combined bounds on the 
gravitino and sgoldstino masses.
\end{abstract}
\end{titlepage}
\setcounter{footnote}{0}
\vskip2truecm
\bc
{\bf 1. Introduction}
\ec

A new fundamental scale close to the weak scale, $G_F^{-1/2} \simeq
300 \gev$, may play a r\^ole in solving the gauge hierarchy problem 
of the Standard Model (SM), and allow for unconventional phenomenology 
at colliders, provided that it can be made compatible with existing 
data. An old--fashioned example along these lines is technicolor, more 
fashionable ones identify the new scale with the string scale or some 
compactification scale. Here we concentrate on a possibility that 
arises in supersymmetric extensions of the SM, when not only the 
supersymmetry-breaking mass splittings $\Delta m^2$, but also the 
supersymmetry-breaking scale $\sqrt{F}$ is close to the weak scale: 
$G_F^{-1/2} \sim \Delta m^2 \simlt \sqrt{F}$. Since in a flat 
space-time $F = \sqrt{3} \, m_{3/2} M_P$, where $m_{3/2}$ is the 
gravitino mass and $M_P = (8 \pi G_N)^{-1/2} \simeq 2.4 \times 
10^{18} \gev$ is the Planck mass, models of this kind are 
characterized by a very light gravitino, with $m_{3/2} \simlt 
10^{-3} \ev$. 

Many aspects of the superlight gravitino phenomenology at colliders,
and in particular gravitino production, either in pairs (tagged by a
photon or a jet) or in association with gauginos, have been discussed 
long ago \cite{slgold} and also more recently \cite{slgnew,bfzph}. A 
very useful tool for these discussions is the supersymmetric equivalence
theorem \cite{seq}, which allows to replace the gravitino with its 
goldstino components, in the effective theory valid at the present 
accelerator energies. However, including the goldstino is not the end 
of the story. If supersymmetry is spontaneously broken but linearly
realized in the language of four-dimensional $N=1$ local quantum 
field theory, then the appropriate effective theory must contain, 
besides the goldstino, also its supersymmetric partners, to be called 
here {\em sgoldstinos}.\footnote{Notice that, in the presence of an 
exact R--parity, as assumed throughout this paper, the goldstino is 
R--odd and the sgoldstinos are R--even.} The simplest possible case 
(as well as the easiest one to reconcile with experimental constraints) 
corresponds to pure F--breaking of supersymmetry, with the goldstino 
and the sgoldstinos belonging to a chiral superfield, singlet under 
the full SM gauge group. 

The effective interactions of sgoldstinos with the SM fields and 
with goldstinos can be characterized by suitable effective couplings. 
In most cases, at the lowest non-trivial order in a supersymmetric
derivative expansion, these couplings are proportional to positive 
powers of supersymmetry-breaking masses, and to negative powers of 
the supersymmetry-breaking scale (or, equivalently, of the gravitino 
mass). Therefore, given the present experimental lower bounds on 
supersymmetry-breaking masses for particles with SM gauge interactions, 
sgoldstino production and decay may become phenomenologically relevant 
at the present collider energies, provided that the sgoldstino masses 
and the supersymmetry-breaking scale are not too large.

Most of the existing studies of sgoldstino phenomenology 
\cite{een}--\cite{lsgap} were strongly influenced by 
the model of ref.~\cite{een}, where the sgoldstinos were  
massless at the classical level, and were assumed to acquire 
very small masses after the inclusion of quantum corrections. 
As a result, collider signals of sgoldstinos were studied 
\cite{br}--\cite{dn} only in the limit of vanishing sgoldstino
masses. However, sgoldstino mass terms are allowed by the
generic structure of supersymmetric effective Lagrangians,
as can be explicitly verified \cite{bfz3}. Moreover, it was 
recently shown \cite{bfzff} that the situation in which the 
sgoldstinos and the gravitino are very light, whilst the 
superpartners of the SM particles are heavy, is generically 
unstable against quantum corrections, i.e. technically unnatural. 
Therefore, a more plausible starting point for discussing sgoldstino 
phenomenology at colliders is to assume that the sgoldstino masses 
are arbitrary parameters, to be constrained only by experiment. 
This is the approach that will be followed in the present paper.

The plan of the paper is as follows. In the rest of this section 
we summarize the assumptions underlying our analysis. These assumptions
are much less restrictive than those of the existing literature, 
but still depart from full generality. Readers who are not familiar 
with the formalism of supersymmetric effective Lagrangians, and are
only interested in the phenomenological aspects of our work, can 
skip this part and move directly to sect.~2. There we give a 
systematic discussion of the most important sgoldstino decay modes, 
focusing on important issues such as the sgoldstino total width 
and branching ratios. In sect.~3 we discuss the mechanisms for 
sgoldstino production in $e^+ e^-$ collisions, with emphasis on
those that are most important for LEP energies. In sect.~4 we 
summarize the resulting signals at LEP and we present our conclusions. 

We conclude this introduction by recalling the assumptions on the
effective theory  underlying our analysis. They may be useful for 
the readers who want to re-derive, starting from the general 
formalism of supersymmetry, the effective couplings used in the 
following sections.

We consider a four-dimensional effective theory with global $N=1$ 
supersymmetry and $SU(3)_C \times SU(2)_L \times U(1)_Y$ gauge
symmetry. The building blocks of such a theory are the fields of 
the Minimal Supersymmetric extension of the Standard Model (MSSM), 
plus a gauge-singlet chiral superfield $Z \equiv (z,\grav,\auxz)$.  

The most general effective Lagrangian with the above field content 
is determined, up to higher-derivative terms, by a gauge kinetic 
function $f$, a superpotential $w$ and a K\"ahler potential $K$
(see, e.g., ref.~\cite{wb}). A detailed discussion of the conditions 
to be imposed on $f$, $w$ and $K$ to obtain a fully realistic 
model will be given elsewhere \cite{bfprz}. Here we give a 
simplified treatment, mentioning only those features that are 
directly relevant for the study of sgoldstino phenomenology.
First, we make the simplifying assumption that the gauge 
kinetic function $f$, which in  principle transforms as 
a symmetric product of adjoint representations, factorizes into 
three independent gauge-invariant functions, one for each factor 
of the gauge group. Then, we assume that the only source of CP 
violation is the standard Kobayashi-Maskawa phase, so that, 
apart from the Yukawa couplings, we can restrict ourselves to real 
parameters and vacuum expectation values (VEVs). As already 
announced, we assume that supersymmetry is spontaneously broken
by $\langle \partial w / \partial z \rangle \equiv F \ne 0$, 
with vanishing VEVs for the auxiliary fields of all the other 
chiral and vector multiplets. This allows us to identify the 
fermionic field $\grav$ with the goldstino. It is not restrictive 
to take $F$ real and positive, so that $\sqrt{F}$ can be 
identified with the supersymmetry-breaking scale. The spin-0 
complex field $z \equiv (S + i P) / \sqrt{2}$ contains then 
the sgoldstinos, one CP--even ($S$) and the other CP--odd ($P$). 
We then assume that the gauge symmetry is spontaneously broken by 
non-vanishing VEVs of the neutral components of the MSSM Higgs 
doublets, $H_1$ and $H_2$. To guarantee that $\rho=1$ at tree level, 
as in the MSSM, we impose a custodial symmetry on K, assuming that 
it depends only on ($H_1 H_2, \ov{H_1 H_2}, H_1^{\dagger} H_1 + 
H_2^{\dagger} H_2$), but not on $(H_1^{\dagger} H_1 - H_2^{\dagger} 
H_2)$. To simplify the discussion further, and to avoid the 
proliferation of parameters, we finally assume that there is no 
sgoldstino-Higgs mixing, and that squarks, sleptons, gluinos, 
charginos, neutralinos and Higgs bosons are sufficiently heavy 
not to play a r\^ole in sgoldstino production and decay. We can 
thus take $S$ and $P$ as mass eigenstates, with eigenvalues $m_S^2$
and $m_P^2$, respectively.  Despite its simplicity, the present 
context will allow us to generalize considerably the existing 
studies on sgoldstinos at colliders \cite{br}--\cite{dn}. A more 
general treatment of the interplay between $SU(2) \times U(1)$ 
and supersymmetry breaking will be given elsewhere \cite{bfprz}.

As explained in \cite{grk}, we can use the freedom to perform 
analytic field redefinitions allowed by gauge invariance, and 
move to a field basis ({\em normal coordinates}) such that, at 
the minimum of the potential, all chiral superfields have 
canonical kinetic terms and, in addition, all the derivatives 
of the K\"ahler potential with respect to one chiral superfield 
and $n>1$ antichiral superfields (or vice-versa) have vanishing 
VEVs. Moreover, this still leaves sufficient freedom to redefine 
$Z$ by a suitable constant shift, to ensure that $\langle z 
\rangle = 0$. In the following we shall always assume normal
coordinates and $\langle z \rangle = 0$: this will lead to simple 
formulae for the mass spectrum and the interactions, with no 
further loss of generality. 

\vspace{1cm}
\bc
{\bf 2. Sgoldstino decays}
\ec

Since we have assumed that squarks, sleptons, gluinos, charginos, 
neutralinos and Higgs bosons are sufficiently heavy to play no 
r\^ole in the decays of the sgoldstinos, we are left with the 
following possibilities for two-body sgoldstino decays:
\be
\label{ztwo}
S \; (P) \; \longrightarrow 
\grav \grav \, ,
\gamma \gamma \, ,
g g  \, , 
\gamma Z \, , 
Z Z \, , 
W^+ W^- \, ,  
f \ov{f} \, ,
\ee
and finally
\be
\label{spp}
S  \longrightarrow P P  \, .
\ee
Three-body decays such as $S \, (P) \, \rightarrow g g g$,  $W^+ 
W^- \gamma$, $W^+ W^- Z$, and four-body decays such as $S \, (P) 
\, \rightarrow g g g g$, $W^+ W^- \gamma \gamma$, $W^+ W^- Z Z$
are also possible. We shall neglect all of them here, since they 
are sub-leading in a perturbative expansion in the gauge coupling 
constants. We now discuss one by one the different decay channels. 
Our notation and conventions are the same as in ref.~\cite{bfz3}. 
For simplicity, here and in the following we shall always assume 
$m_S, m_P \gg 1 \gev$, to avoid all theoretical subtleties connected 
with the non--perturbative aspects of the strong interactions: 
sgoldstinos with masses up to a few GeV would deserve a dedicated 
phenomenological study.

\vspace{0.7cm}
\begin{center}
$S \;  (P) \longrightarrow \grav \grav$
\end{center}
These decays are controlled by the effective interactions
\be
\cl_{z\grav\grav} =  
- {1 \over 2 \sqrt{2}} {m_S^2 \over F} S \, \grav \grav 
+ {i \over 2 \sqrt{2}} {m_P^2 \over F} P \, \grav \grav
+ {\rm h.c.}  \, ,
\label{lsggogo}
\ee
which give \cite{bfz3}
\be
\Gamma ( S \longrightarrow \grav \grav ) = {m_S^5 \over
32 \pi F^2} \, ,
\;\;\;\;\;
\Gamma ( P \longrightarrow \grav \grav ) = {m_P^5 \over
32 \pi F^2} \, .
\ee 

\vspace{0.7cm}
\begin{center}
$S \;  (P) \longrightarrow \gamma \gamma$
\end{center}
The relevant terms in the effective Lagrangian are 
\be
\cl_{z\gamma\gamma} =   
- {1 \over 2 \sqrt{2}} {M_{\gamma \gamma} \over F} S 
\, F^{\mu \nu} F_{\mu \nu} 
+ {1 \over 4 \sqrt{2}} {M_{\gamma \gamma} \over F} P \,  
\epsilon^{\mu \nu \rho \sigma} F_{\mu \nu} F_{\rho \sigma}
\, ,
\label{lsgphph}
\ee
where $F_{\mu \nu} = \partial_{\mu} A_{\nu} - \partial_{\nu} 
A_{\mu}$ is the electromagnetic field strength, and
\be
\label{mgg}
\mgg = M_1 \cos^2 \theta_W +  M_2 \sin^2 \theta_W \, .
\ee
In the above equation, $M_1$ and $M_2$ are the diagonal mass 
terms for the $U(1)_Y$ and $SU(2)_L$ gauginos, respectively. From 
eq.~(\ref{lsgphph}) we can easily compute the decay rates, 
generalizing the results of \cite{bfz3}:
\be
\Gamma ( S \longrightarrow \gamma \gamma) = {m_S^3 M^2_{\gamma
\gamma} \over 32 \pi F^2} \, ,
\;\;\;\;\;
\Gamma ( P \longrightarrow \gamma \gamma ) = {m_P^3 M^2_{\gamma
\gamma} \over 32 \pi F^2} \, ,
\ee 

\vspace{0.7cm}
\begin{center}
$S \;  (P) \longrightarrow g g $
\end{center}
The discussion of these decay modes is a straightforward 
generalization of the previous ones to the non-Abelian 
case. They are controlled by the effective interactions
\be
\cl_{zgg} =  
- {1 \over 2 \sqrt{2}} {M_3 \over F} S \, 
G^{\mu \nu \; \alpha} G_{\mu \nu}^{\alpha} 
+ {1 \over 4 \sqrt{2}} {M_3 \over F} P \,  
\epsilon^{\mu \nu \rho \sigma} G_{\mu \nu}^{\alpha} 
G_{\rho \sigma}^{\alpha} \, ,
\label{lsggg}
\ee
where $G_{\mu \nu}^{\alpha}$ is the $SU(3)$ field strength and 
$M_3$ is the gluino mass. From eq.~(\ref{lsggg}) we obtain
\be
\Gamma ( S \longrightarrow g g) = {m_S^3 M_3^2 \over
4 \pi F^2} \, ,
\;\;\;\;\;
\Gamma ( P \longrightarrow g g  ) = {m_P^3 M_3^2 \over
4 \pi F^2} \, .
\ee 

\vspace{0.5cm}
\begin{center}
$S \;  (P) \longrightarrow  \gamma Z $
\end{center}
Since the gaugino block of the neutralino mass matrix and the 
gauge boson mass matrix cannot be simultaneously diagonalized
(apart from the special case $M_2 = M_1$), these decay modes 
may be phenomenologically relevant and cannot be neglected.
They are controlled by the effective interactions \cite{bfprz}
\be
\cl_{z \gamma Z}  =  
- {1 \over \sqrt{2}} {M_{\gamma Z} \over F} S 
\, F^{\mu \nu} Z_{\mu \nu} 
+ {1 \over 2 \sqrt{2}} {M_{\gamma Z} \over F} P \,  
\epsilon^{\mu \nu \rho \sigma} F_{\mu \nu} Z_{\rho \sigma}
\, ,
\label{lsgphz}
\ee
where $Z_{\mu \nu} = \partial_{\mu} Z_{\nu} - \partial_{\nu} Z_{\mu}$ 
is the Abelian field strength for the $Z$ boson, and 
\be
\label{mgz}
\mgz = (M_2 - M_1) \sin \theta_W \cos \theta_W \, .
\ee From eq.~(\ref{lsgphz}) we can easily compute the decay rates
\be
\Gamma ( S \longrightarrow \gamma Z) = 
\displaystyle{ \mgz^2 m_S^3 \over 16 \pi F^2}
\left( 1 - {m_Z^2 \over m_S^2} \right)^3  \, ,
\;\;\;\;\;
\Gamma ( P \longrightarrow \gamma Z) = 
\displaystyle{ \mgz^2 m_P^3 \over 16 \pi F^2}
\left( 1 - {m_Z^2 \over m_P^2} \right)^3  \, .
\ee

\vspace{0.5cm}
\begin{center}
$S \;  (P) \longrightarrow  Z Z $
\end{center}
The discussion of these decay modes is complicated by the fact that 
the corresponding interactions originate not only from the generalized
kinetic terms for the electroweak gauge bosons, but also, in the case
of $S$, from the generalized kinetic terms of the Higgs bosons. These 
decay modes are controlled by the effective interactions \cite{bfprz}
\be
\cl_{zZZ} = 
- {1 \over 2 \sqrt{2}} {M_{ZZ} \over F} S 
\, Z^{\mu \nu} Z_{\mu \nu} 
-{m_Z^2 \mu_a \over \sqrt{2} F} S Z^{\mu} Z_{\mu}
+ {1 \over 4 \sqrt{2}} {M_{ZZ} \over F} P \,  
\epsilon^{\mu \nu \rho \sigma} Z_{\mu \nu} Z_{\rho \sigma}
\, ,
\label{lsgzz}
\ee
where 
\be
\label{mzz}
\mzz = M_1 \sin^2 \theta_W + M_2 \cos^2 \theta_W \, ,
\ee
and $\mu_a$ is a diagonal mass term for the antisymmetric 
neutralino combination, $(\tilde{H}_1^0 - \tilde{H}_2^0)/
\sqrt{2}$, analogous (but not identical) to the so-called 
$\mu$-term of the MSSM. From eq.~(\ref{lsgzz}) we find
\bea
\Gamma ( S \longrightarrow ZZ ) & = & 
{1 \over 32 \pi F^2 m_S} \left[ \mzz^2 \left(m_S^4-4m_S^2 m_Z^2
+6 m_Z^4\right)\right.\nn\\
&-&12 \mzz \mu_a m_Z^2 \left({m_S^2 \over 2}-m_Z^2 \right)\nn\\
&+&\left. 2 \mu_a^2 m_Z^4 \left({m_S^4 \over 4 m_Z^4} 
- {m_S^2 \over m_Z^2} +3\right)\right]\sqrt{1-{4 m_Z^2 \over m_S^2}}
\, ,
\eea
\be
\Gamma ( P \longrightarrow ZZ ) = 
{\mzz^2 m_P^3 \over 32 \pi F^2}
\left( 1-{4 m_Z^2 \over m_P^2} \right)^{3/2}
\, .
\ee

\vspace{0.5cm}
\begin{center}
$S \;  (P) \longrightarrow W^+ W^-$
\end{center}
The discussion of these decay modes is the obvious generalization 
of the previous ones. They are controlled by the effective 
interactions \cite{bfprz}
\be
\cl_{zWW} = 
- {1 \over \sqrt{2}} {M_2 \over F} S 
\, W^{\mu \nu \, +} W_{\mu \nu}^- 
-{\sqrt{2} m_W^2 \mu_a \over F} S W^{\mu \, +} W_{\mu}^-
+ {1 \over 2 \sqrt{2}} {M_2 \over F} P \,  
\epsilon^{\mu \nu \rho \sigma} W_{\mu \nu}^+ W_{\rho \sigma}^-
\, ,
\label{lsgww}
\ee
where  $W_{\mu \nu}^{\pm} = \partial_{\mu} W_{\nu}^{\pm}
- \partial_{\nu} W_{\mu}^{\pm}$ are the Abelian field 
strengths for the $W^\pm$ bosons, and $\mu_a$, already defined 
above, can be also identified with the diagonal higgsino entry 
in the chargino mass matrix. From eq.~(\ref{lsgww}) we can 
easily compute the decay rates
\bea
\Gamma ( S \longrightarrow W^+ W^- ) & = & 
{1 \over 16 \pi F^2 m_S} \left[ M_2^2 \left(m_S^4-4m_S^2 m_W^2
+6 m_W^4\right)\right.\nn\\
&-&12 M_2 \mu_a m_W^2 \left({m_S^2 \over 2}-m_W^2 \right)\nn\\
&+&\left.  2 \mu_a^2 m_W^4  \left({m_S^4 \over 4 m_W^4} 
- {m_S^2 \over m_W^2} +3\right)\right]\sqrt{1-{4 m_W^2 \over m_S^2}}
\, ,
\eea
\be
\Gamma ( P \longrightarrow W^+ W^- ) = 
{M_2^2 m_P^3 \over 16 \pi F^2}
\left(1-{4 m_W^2 \over m_P^2} \right)^{3/2} 
\, .
\ee

\vspace{0.5cm}
\begin{center}
$S \;  (P) \longrightarrow f \ov{f} $
\end{center}
As discussed in \cite{bpz}, the Yukawa couplings of sgoldstinos
to fermions are expected to be suppressed by a factor $m_f/
\sqrt{F}$, where $m_f$ is the fermion mass. This can be justified
in terms of the same chiral symmetry that suppresses the 
off-diagonal (left-right) contributions to the sfermion mass 
matrices. We then expect these decay modes to be important
only for very heavy sgoldstinos decaying into top-antitop 
pairs, thus we shall always neglect them in the following.

\vspace{0.5cm}
\begin{center}
$S \longrightarrow P P $
\end{center}

If $m_S > 2 m_P$, this decay is kinematically allowed and must be 
taken into account. At the level of the effective theory, the 
corresponding cubic coupling is a free parameter, unrelated with 
the spectrum. In the following we shall always neglect this decay mode,
consistently with the strategy of focusing the attention on the
lighter sgoldstino, the most likely to be discovered first.

\vspace{0.5cm}
Now that we have explicit formulae for the most important two-body 
decays of the sgoldstinos, we can move to the discussion of their
total widths and branching ratios. Since the expressions for the 
partial widths of $S$ and $P$ are identical in all cases of practical 
interest, we shall give such a discussion only in the case of $S$.

The parameters controlling $S$ decays are $m_S$, $\sqrt{F}$, 
the gaugino masses $(M_3,M_2,M_1)$ and the higgsino mass $\mu_a$.
In the following, when giving numerical examples, we shall focus 
our attention on the dependences on $m_S$ and $\sqrt{F}$, by 
making for the remaining parameters the two representative 
choices given in table~1. For most values of $\sqrt{F}$ to be
considered in the following, these choices should be comfortably
compatible with the present experimental limits on R-odd
supersymmetric particles, coming from LEP and Tevatron searches.
\begin{table}[ht]
\begin{center}
\begin{tabular}{|c|c|c|c|c|}
\hline
& $M_3$ & $M_2$ & $M_1$ & $\mu_a$ \\
\hline
(a) & 400 & 300 & 200 & 300 \\
\hline
(b) & 350 & 350 & 350 & 350 \\
\hline
\end{tabular}
\end{center}
\caption{Two representative choices for the gaugino and 
higgsino mass parameters affecting sgoldstino decays.
All masses are expressed in GeV.}
\end{table}

Since all the two-body decay widths are proportional to 
$F^{-2}$, the dependence on $F$ drops out of the discussion of 
the $S$ branching ratios. The latter are shown in fig.~1, as
functions of $m_S$, for the two parameter choices of table~1.
\begin{figure}[htbp]
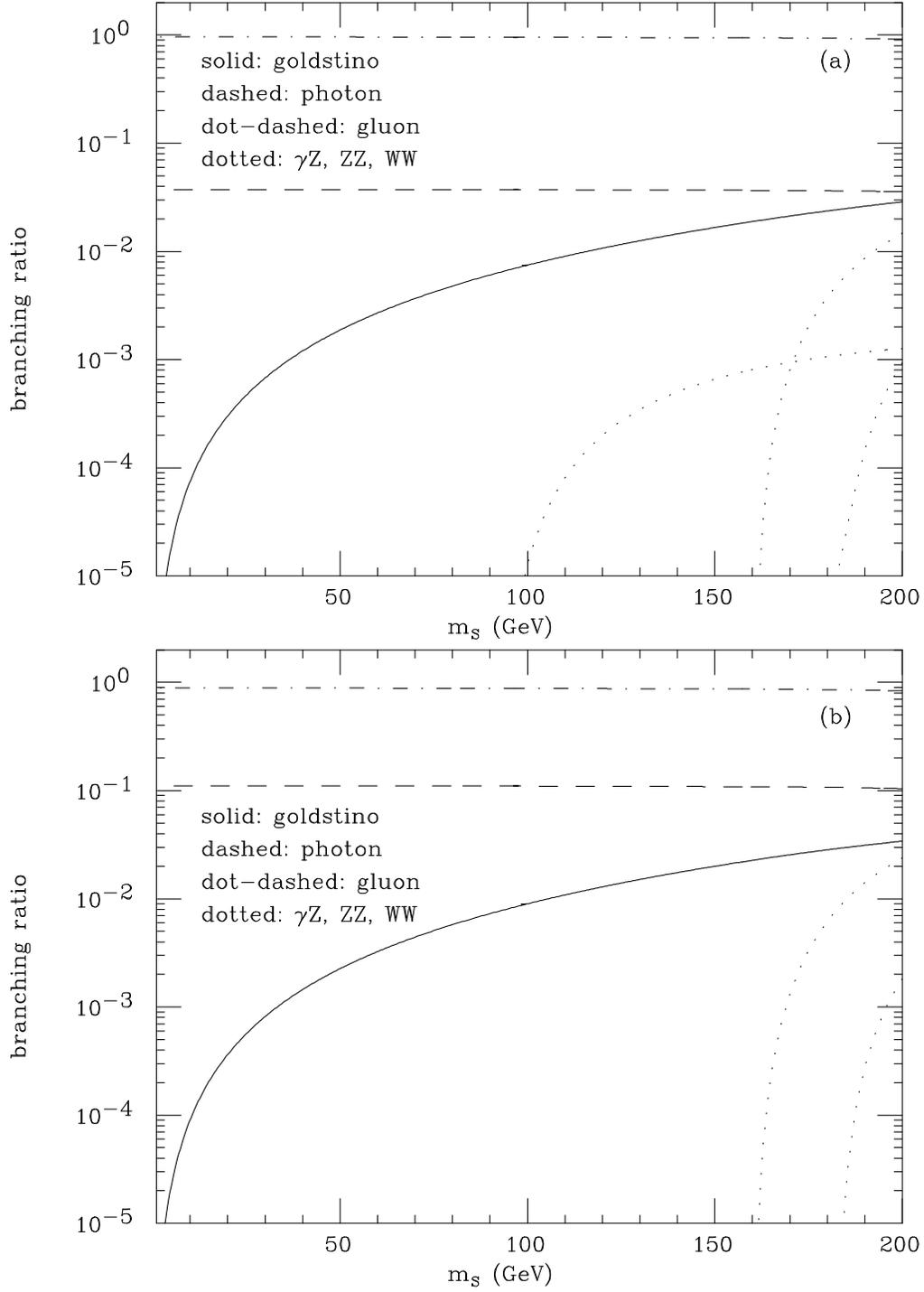

\begin{center}
\epsfig{figure=prz1a.ps,height=9.5cm,angle=0}
\epsfig{figure=prz1b.ps,height=9.5cm,angle=0}
\end{center}
\caption{{\it The $S$ branching ratios, as functions 
of $m_S$, for the parameter choices of table~1.}}
\label{brs}
\end{figure}
We can see that the dominant decay mode is always the one
into gluons. Even in the extreme case (b), this decay 
mode dominates over the one into photons, because of the 
color factor $8$. Decays into goldstinos can become important 
only when $m_S$ is of the order of $M_3$ or larger: in this case, 
however, we would expect gauginos (produced in pairs or in 
association with a gravitino) to be detected before sgoldstinos. 

The other important quantity is the total $S$ width, $\Gamma_S$,
controlled by the ratios between the relevant mass parameters and 
the supersymmetry-breaking scale. Small values of these ratios 
correspond to relatively long-lived sgoldstinos, large values 
of these ratios correspond to broad, strongly coupled sgoldstinos: 
to keep the particle interpretation and the validity of our 
approximations, we must require, among the other things, that 
$\Gamma_S/m_S \ll 1$. We show in fig.~2 contours corresponding to 
constant values of $\Gamma_S/m_S$ in the $(m_S,\sqrt{F})$-plane, 
for the parameter choices of table~1. 
\begin{figure}[htbp]
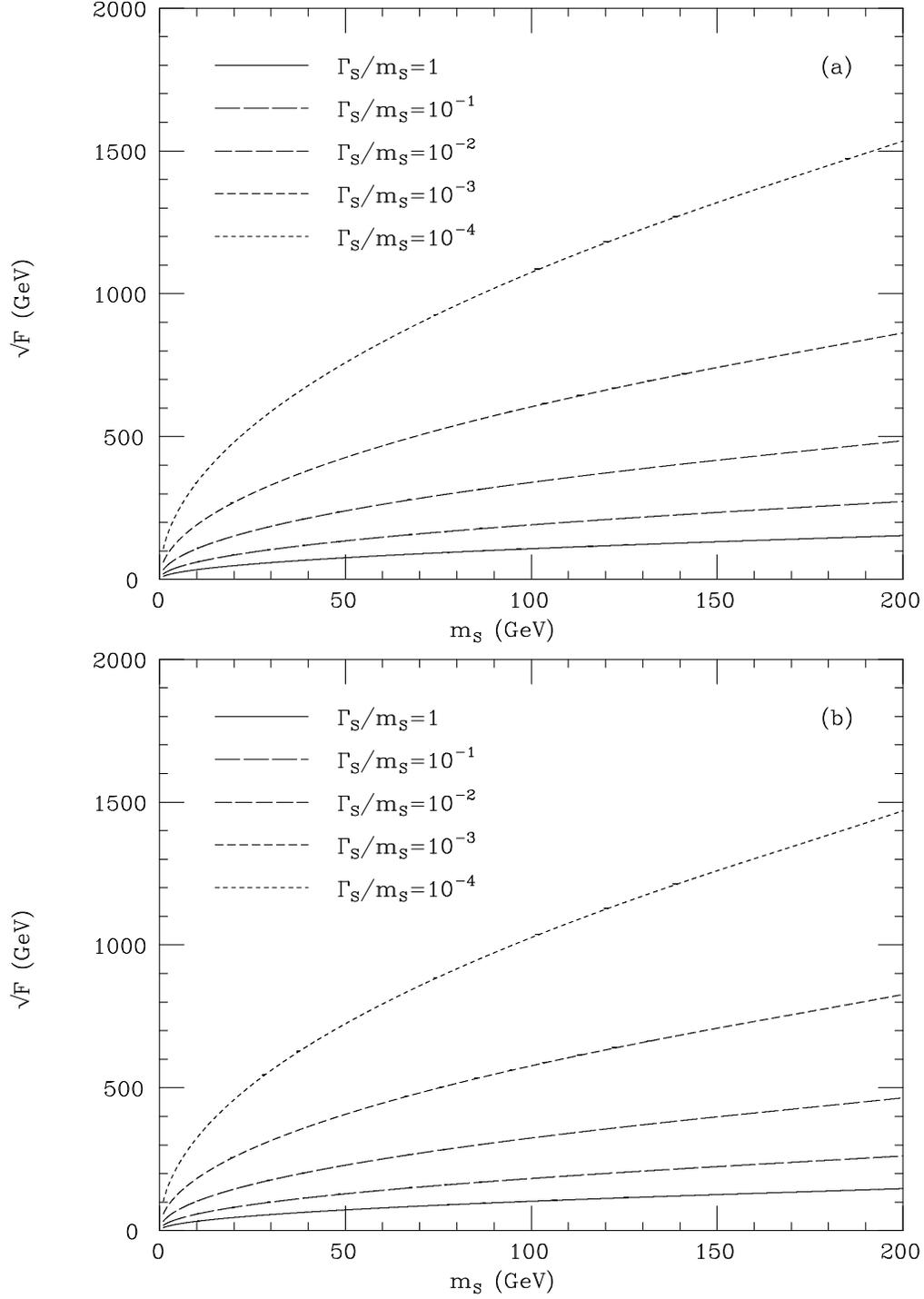

\begin{center}
\epsfig{figure=prz2a.ps,height=9.5cm,angle=0}
\epsfig{figure=prz2b.ps,height=9.5cm,angle=0}
\end{center}
\caption{{\it Lines corresponding to fixed values of 
$\Gamma_S/m_S$ in the $(m_S,\sqrt{F})$ plane, for the 
parameter choices of table~1.}}
\label{wis}
\end{figure}
As we shall see in sects.~3 and 4, the region of parameter space
of present experimental interest is such that sgoldstinos can always 
be treated as very narrow resonances.

A question that should be asked is whether the sgoldstino 
widths can be so small that sgoldstinos can travel for an 
experimentally significant length in a detector before 
decaying. Since, as we have seen, the dominant decay mode
is always the one into gluons, the typical distance traveled
by a sgoldstino $S$ of mass $m_S$ and energy $E_S$ can be 
written as
\be
\label{decl}
L \simeq
\left( {\sqrt{F} \over 1 \tev} \right)^4
\left( {1 \gev \over m_S} \right)^3
\left( {300 \gev \over M_3} \right)^2
\sqrt{{E_S^2 \over m_S^2} - 1}
\cdot \left( 2.75 \times 10^{-2} {\, \mu m} \right)  \, .
\ee   
Again, we shall check in sects.~3 and 4 that, for $m_S \gg 1 \gev$, 
and values of the other parameters not excluded by the present data 
but leading to non-negligible production cross-sections at LEP, 
sgoldstinos always decay within a $\mu m$ from the primary interaction 
vertex.

Before concluding this section, we should also mention the 
possibility of three-body decays such as $S \to P \, f \,
\ov{f}$ (or $P \to S \, f \, \ov{f}$), where $f$ is a light
matter fermion, induced by local four-point interactions
that are not controlled by the gauge couplings, of the form
\be
\cl_{f\ov{f}zz} = 
{1 \over 2 F^2} 
\left( 
\tilde{m}^2_f \ov{f} \ov{\sigma}^\mu f 
+
\tilde{m}^2_{f^c} \ov{f^c} \ov{\sigma}^\mu f^c
\right)
\left( S \partial_\mu P - P \partial_\mu S \right) 
\, ,
\label{lsgsgff}
\ee
where $\tilde{m}^2_f$ and $\tilde{m}^2_{f^c}$ are 
the diagonal supersymmetry-breaking masses for the
left- and right-handed sfermions, respectively.
The corresponding widths are easily calculated
in the limit of massless fermions:
\be
\label{g3body}
\Gamma(S \to P \, f \, \ov{f}) = 
{N_f (\tilde{m}^4_f + \tilde{m}^4_{f^c}) \over 6144 
\pi^3 m_S^2 F^4} \left[ {m_S^8 - 8 m_S^6 m_P^2 
+ 8 m_S^2 m_P^6 - m_P^8 \over m_S} + 12
m_P^4 m_S^3 \log {m_S^2 \over m_P^2} \right] \, ,
\ee
where $N_f=1$ for leptons and $N_f=3$ for quarks, 
and similarly for $P \to S \, f \, \ov{f}$. 
Equation~(\ref{g3body}) simplifies considerably when the 
mass of the sgoldstino in the final state can be neglected:
\be
\Gamma(S \to P \, f \, \ov{f}) = 
{N_f (\tilde{m}^4_f + \tilde{m}^4_{f^c}) m_S^5 
\over 6144 \pi^3 F^4} \, ,
\;\;\;\;\; (m_P=0) \, .
\ee
Notice, however, that these decays are strongly suppressed 
not only by the phase space, but also by a higher power
of the supersymmetry-breaking scale at the denominator.
Moreover, this decay mode is of course relevant only for the 
heavier sgoldstino, thus, on the same basis as for $S \to PP$, 
we could safely neglect it in the previous discussion.

\vspace{1cm}
\bc
{\bf 3. Sgoldstino production}
\ec

We now review the most important mechanisms for sgoldstino
production at $e^+e^-$ colliders, and especially at LEP. As 
before, whenever the formulae for $S$ and $P$ are identical
in form, apart from the obvious substitution $S \leftrightarrow
P$, we refer to $S$ only.

Since the sgoldstino couplings to light fermions are suppressed
by the corresponding fermion masses (as it is the case for the SM
Higgs), resonant production in the $s$-channel can be neglected.

At LEPI, we can consider the possibility of $Z \to S \, \gamma$
decays, whose rate can be easily calculated from the effective 
Lagrangian of eq.~(\ref{lsgphz}):
\be
\label{zsg}
\Gamma(Z \longrightarrow S \, \gamma) = 
{\mgz^2 m_Z^3 \over 48 \pi F^2} 
\left( 1 - {m_S^2 \over m_Z^2} \right)^3 
\, .
\ee 
To give a measure of the LEPI sensitivity, we show in fig.~3 
contours of constant branching ratios, $BR(Z \to S \, \gamma) 
\equiv \Gamma(Z \to S \, \gamma)/\Gamma_Z$, in the $(m_S,\sqrt{F})$ 
plane, for the parameter choice (a) of table~1 [the parameter
choice (b) leads to a vanishing effective coupling, $\mgz=0$].
\begin{figure}[htbp]
\begin{center}
\epsfig{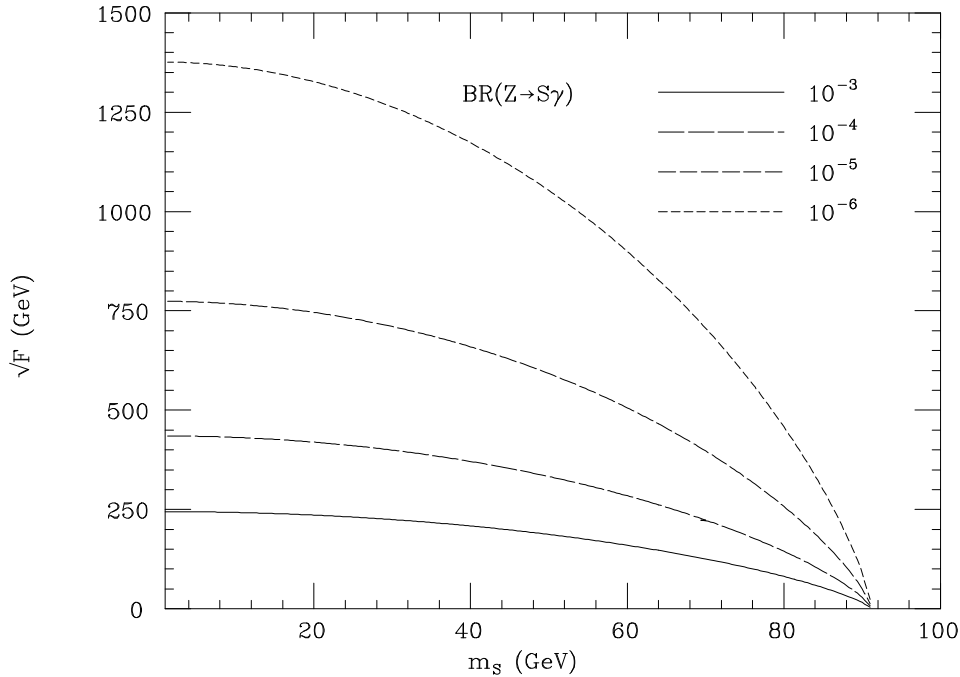}
\end{center}
\caption{{\it Lines corresponding to fixed values of 
$BR(Z \to S \, \gamma)$, in the $(m_S,\sqrt{F})$ plane, 
for the parameter choice (a) of table~1.}}
\label{brzsg}
\end{figure}

More generally, we can consider the process $e^+ e^- \to S \gamma$. 
At the classical level, there are only two Feynman diagrams to compute, 
corresponding to s-channel $\gamma$ and $Z$ exchange. Neglecting both
the electron mass and the $Z$ width, the differential cross-section is
\be
\label{xsg}
{d\sigma \over d \cos \theta} 
\left(e^+ e^- \rightarrow S \gamma \right) = 
{|\Sigma|^2 s \over 64 \pi F^2} 
\left(1 - {m_S^2 \over s} \right)^3  
\left(1 + \cos^2 \theta \right) \, , 
\ee
where
\be
|\Sigma|^2 = { e^2  Q_e^2  \mgg^2  \over 2 s }
+
{ g_Z^2 (v_e^2 + a_e^2 ) \mgz^2  s \over 2 (s-m_Z^2)^2}
+
{e Q_e g_Z v_e \mgg \mgz \over (s-m_Z^2)} \, ,
\ee
$v_e=T_{3e}/2-Q_e \sin^2 \theta_W$, $a_e=-T_{3e}/2$, $T_{3e}=-1/2$, 
$Q_e=-1$, $g_Z=e/(\sin \theta_W \cos \theta_W)$, and $\theta$ is 
the scattering angle in the center-of-mass frame. In the limit 
$m_S=0$, $M_1=M_2$ (i.e. $\mgz=0$), we recover the results of 
ref.~\cite{dr}. This process is particularly relevant at LEPII. To 
give a measure of the LEPII sensitivity, we draw in fig.~4 contours 
of constant $\sigma(e^+ e^- \to S \, \gamma)$ in the $(m_S,\sqrt{F})$ 
plane, for $\sqrt{s} = 200 \gev$ and the two parameter choices of 
table~1.
\begin{figure}[htbp]
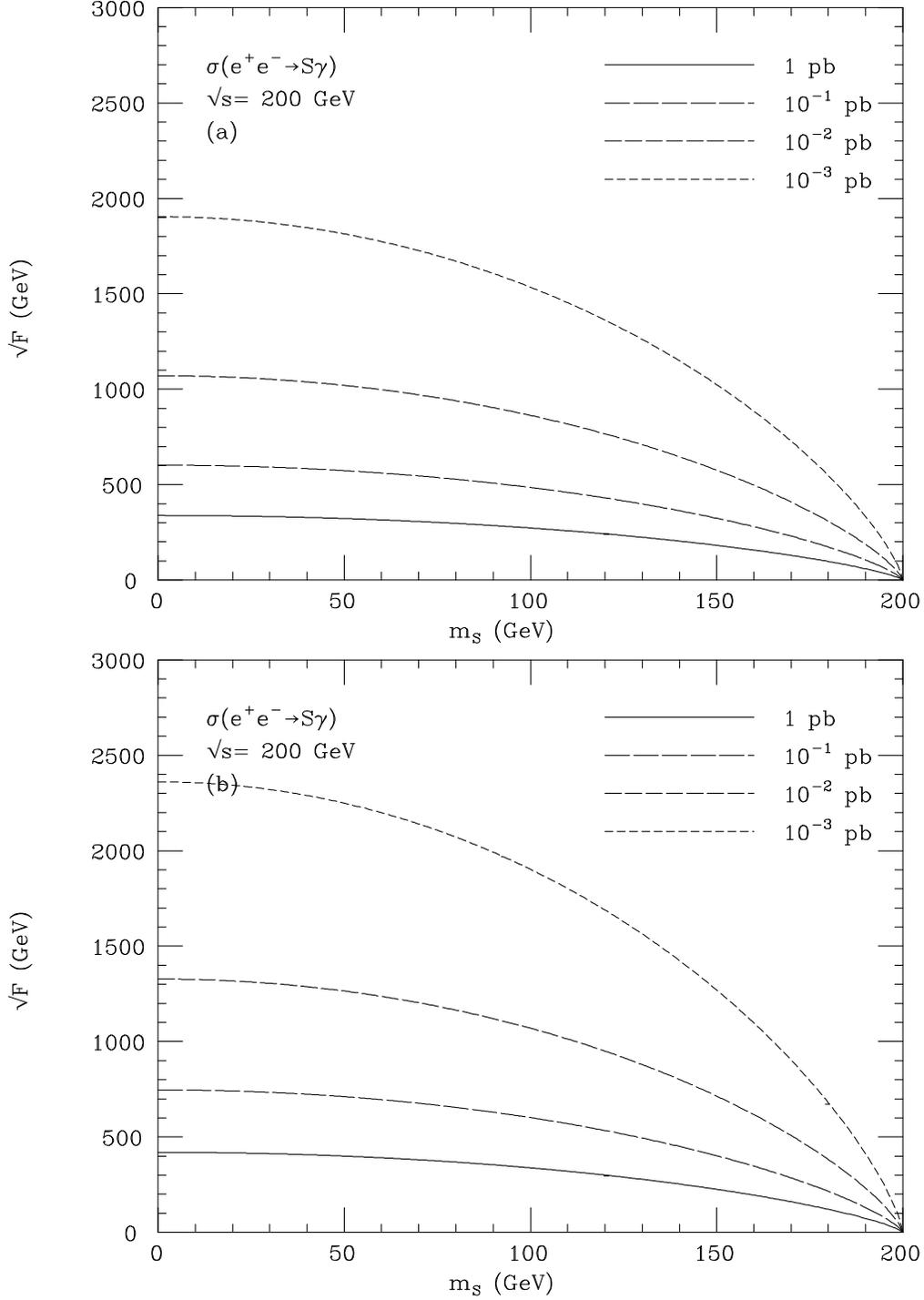

\begin{center}
\epsfig{figure=prz4a.ps,height=9.5cm,angle=0}
\epsfig{figure=prz4b.ps,height=9.5cm,angle=0}
\end{center}
\caption{{\it Lines corresponding to fixed values of $\sigma
(e^+ e^- \to S \, \gamma)$ for $\sqrt{s}=200 \gev$, in the 
$(m_S,\sqrt{F})$ plane and for the parameter choices of table~1.}}
\label{siggsg}
\end{figure}

Another process to be considered is $e^+ e^- \to P Z$, an obvious 
generalization of $e^+ e^- \to P \gamma$. The differential 
cross-section is given by
\be
\label{xpz}
{d\sigma \over d \cos \theta} 
\left( e^+ e^- \rightarrow P Z \right) = 
{|\Sigma_P|^2 \over 32 \pi s^2 F^2} 
\sqrt{(s - m_P^2 - m_Z^2)^2 - 4 m_P^2 m_Z^2} 
\, , 
\ee
where
\be
|\Sigma_P|^2 =  
\left(
{ e^2  Q_e^2  \mgz^2  \over 2 s }
+
{ g_Z^2 (v_e^2 + a_e^2 ) \mzz^2  s \over 2 (s-m_Z^2)^2}
+
{e Q_e g_Z v_e \mgz \mzz \over (s-m_Z^2)}
\right)
\left( 
t^2 + u^2 - 2 m_P^2 m_Z^2 
\right) \, ,
\ee
and $t$ and $u$ are the Mandelstam variables. The cross-section 
for $e^+ e^- \to S Z$ has some additional complications, because of
the couplings controlled by the higgsino mass parameter $\mu_a$:
\be
\label{xsz}
{d\sigma \over d \cos \theta} 
\left(e^+ e^- \rightarrow S Z \right) = 
{|\Sigma_S|^2 \over 32 \pi s^2 F^2} 
\sqrt{(s - m_S^2 - m_Z^2)^2 - 4 m_S^2 m_Z^2} 
\, , 
\ee
where
\bea
|\Sigma_S|^2 &  = &   
\left[
{ e^2  Q_e^2  \mgz^2  \over 2 s }
+
{ g_Z^2 (v_e^2 + a_e^2 ) \mzz^2  s \over 2 (s-m_Z^2)^2}
+
{e Q_e g_Z v_e \mgz \mzz \over s-m_Z^2}
\right] \!
\left[ t^2 \! + \! u^2 \! + 2 m_Z^2 (2s-m_S^2) \right]
\nonumber \\ 
& + & 
{g_Z^2 \mu_a^2 m_Z^4 (v_e^2 + a_e^2 ) \over (s-m_Z^2)^2}
\left(2s -m_S^2 + {t u \over m_Z^2} \right)
\nonumber \\
& + &
{g_Z \mu_a m_Z^2 \over s - m_Z^2}
\left[ 
{g_Z (v_e^2 + a_e^2 ) \mzz \over s-m_Z^2 }
+
{e Q_e \mgz v_e \over s} 
\right]
\left[ 2 s (s + m_Z^2 -m_S^2) \right] \, .
\eea
We draw in figs.~5 and 6 contours of constant $\sigma(e^+ e^- \to 
P \, Z)$ and $\sigma(e^+ e^- \to S \, Z)$, in the $(m_P,\sqrt{F})$
and in the $(m_S,\sqrt{F})$ plane, respectively, for $\sqrt{s} = 
200 \gev$ and the two parameter choices of table~1.
\begin{figure}[htbp]
\bc
\epsfig{figure=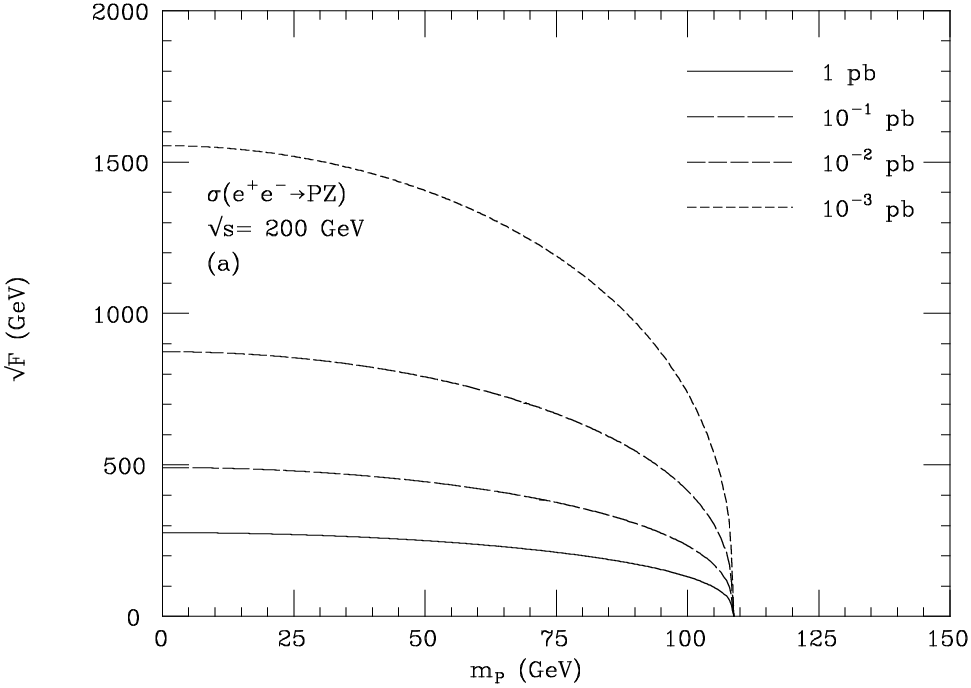,height=9.5cm,angle=0}
\epsfig{figure=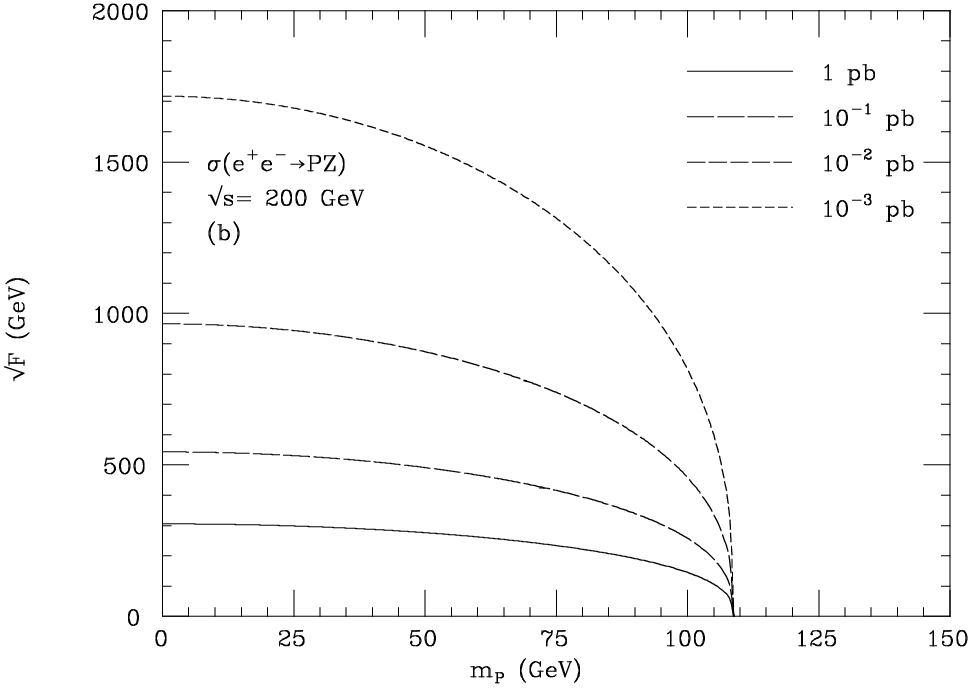,height=9.5cm,angle=0}
\ec
\caption{{\it Lines corresponding to fixed values of $\sigma
(e^+ e^- \to P \, Z)$ for $\sqrt{s}=200 \gev$, in the $(m_P,
\sqrt{F})$ plane and for the parameter choices of table~1.}}
\label{sigzp}
\end{figure}
\begin{figure}[htbp]
\bc
\epsfig{figure=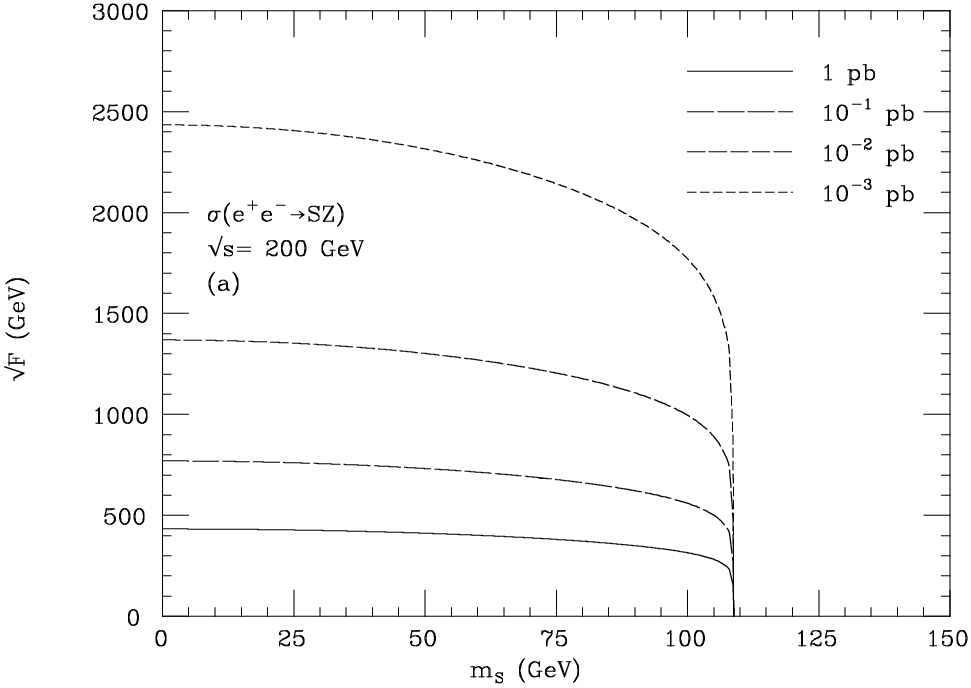,height=9.5cm,angle=0}
\epsfig{figure=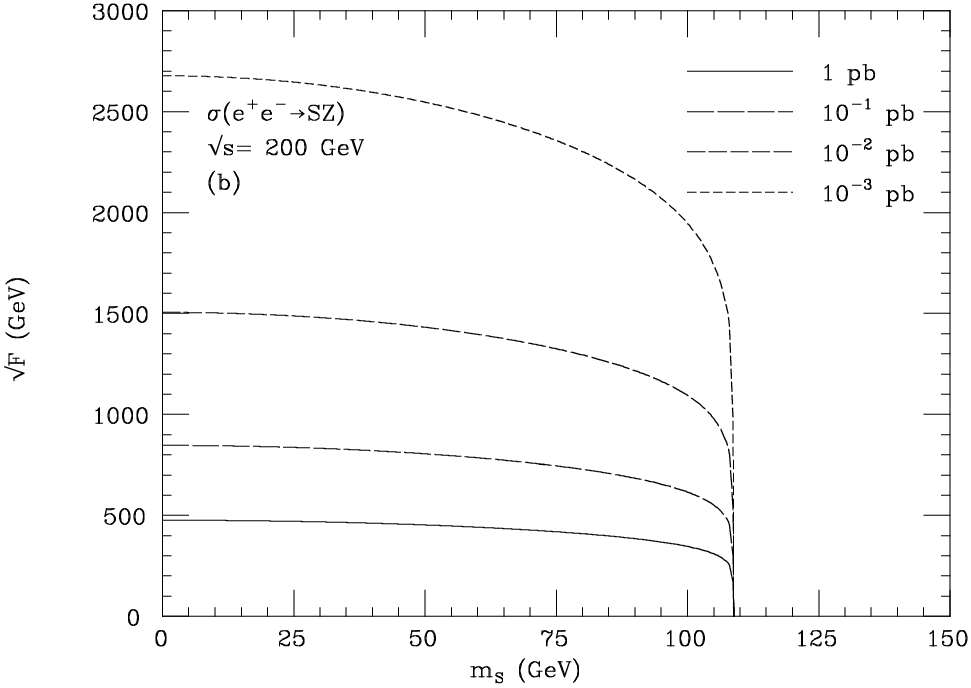,height=9.5cm,angle=0}
\ec
\caption{{\it Lines corresponding to fixed values of $\sigma
(e^+ e^- \to S \, Z)$ for $\sqrt{s}=200 \gev$, in the $(m_S,
\sqrt{F})$ plane and for the parameter choices of table~1.}}
\label{sigzs}
\end{figure}

Other interesting processes at LEPII are $e^+ e^- \to e^+ e^- S$,
occurring via $\gamma \gamma$--, $\gamma Z$-- and $ZZ$--fusion, and
$e^+ e^- \to \nu_e \ov{\nu}_e S$, occurring via $WW$--fusion. We
discuss here only the first process, considering only the $\gamma 
\gamma$--fusion diagram, since at LEP energies all the other diagrams 
give much smaller contributions to the total cross-section, and the 
interference with $e^+ e^- \to Z^{(*)} S \to e^+ e^- S$ is negligible. 
The production of sgoldstinos via $\gamma \gamma$--fusion can be 
described in the Weizs\"acker-Williams approximation, i.e. neglecting 
contributions of off-shell (non-collinear) photons. So doing, the 
cross--section for the process $e^+e^- \to e^+e^- S$ can be expressed 
in terms of the cross--section for the subprocess $\gamma\gamma \to 
S$, where each photon is taken on its mass shell, and carries a 
fraction of the incoming electron momentum which is distributed 
according to the Weizs\"acker-Williams function. In formulae
\cite{wewi}:
\be
\label{ww}
\sigma(s)=\int_{\tau_S}^1 dx_1 \int_{\tau_S/x_1}^1 \!\!\! dx_2 \; 
f^{WW}(x_1) \, f^{WW}(x_2) \, d\sigma_{\gamma\gamma}(x_1 x_2 s) \, ,
\ee
where $s$ is the center-of-mass squared energy, $\tau_S=m_S^2/s$ and
$x_1,x_2$ are the fractions of the incoming
electron and positron momenta carried by the colliding photons.
The Weizs\"acker-Williams distribution function is given by
\be
f^{WW}(x)=\frac{\alpha_{em}}{2\pi}\left[
2 m_e^2 x\left(\frac{1}{m_S^2}-\frac{1-x}{m_e^2 x^2}\right)
+
\frac{1+(1-x)^2}{x}\log\frac{Q^2(1-x)}{m_e^2 x^2}\right]
\ee
where $Q$ is an energy scale of the order of $m_S$. Corrections 
to eq.~(\ref{ww}) are suppressed with respect to $\log(m_e/m_S)$ 
by powers of $m_e/m_S$ (including the zeroth power). From
the effective Lagrangian of eq.~(\ref{lsgphph}), we find
\be
\sigma_{\gamma \gamma}(\hat{s}) = {\mgg^2 m_S^2 \pi \over 4 F^2}
\delta (\hat{s} - m_S^2) \, ,
\ee
from which we deduce, after some trivial manipulations,
\be
\label{sees}
\sigma(e^+ e^- \to e^+ e^- S) = {\mgg^2 \tau_S \pi \over 4 F^2}
\int_{\tau_S}^1 \frac{dx}{x} f^{WW}(x) f^{WW}(\tau_S/x)
\, .
\ee
Numerical results are given in fig.~7, which shows contours of 
constant $\sigma(e^+ e^- \to e^+ e^- S )$ in the $(m_S,\sqrt{F})$ 
plane, for $\sqrt{s} = 200 \gev$ and the two parameter choices of 
table~1.
\begin{figure}[htbp]
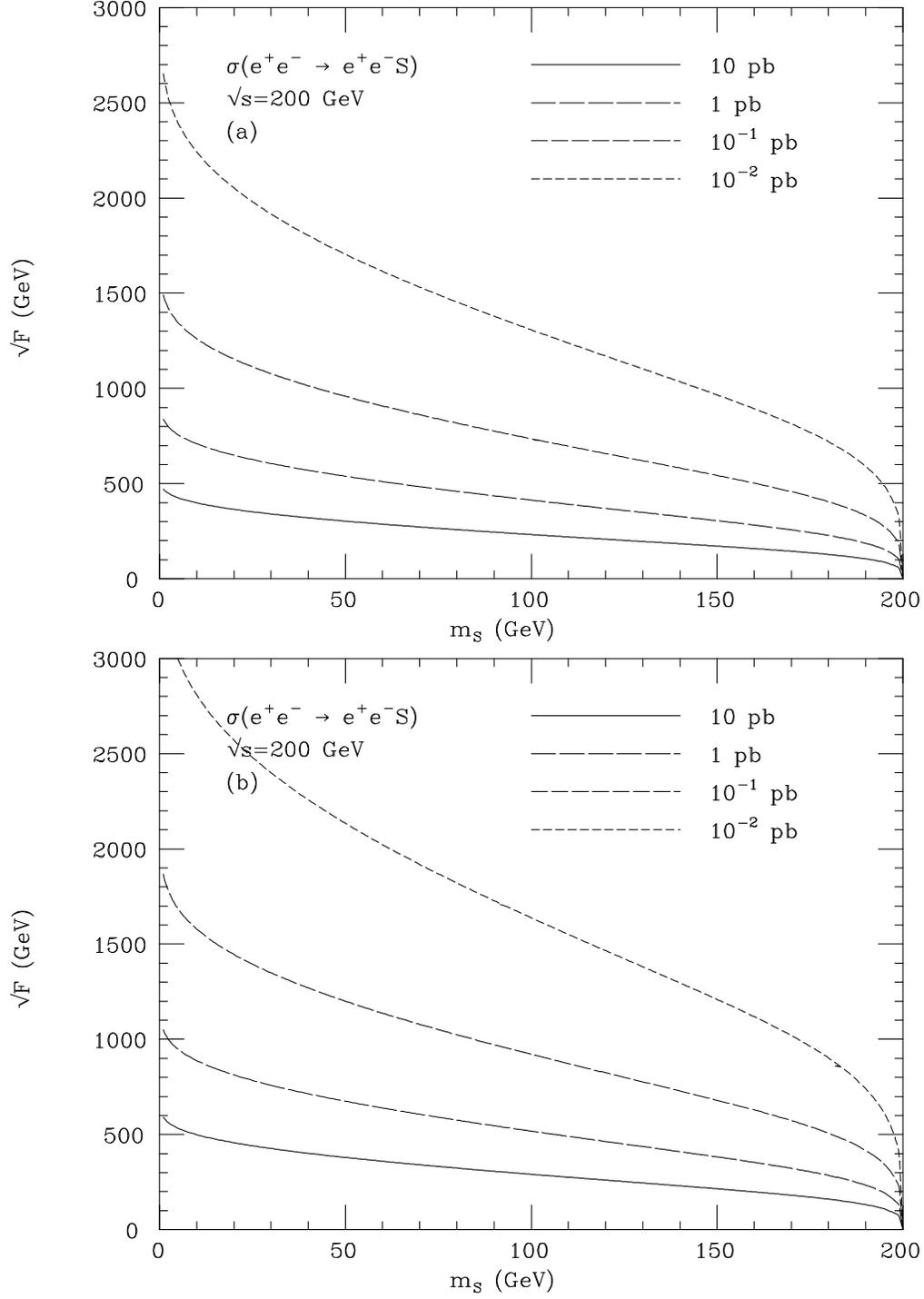

\begin{center}
\epsfig{figure=prz7a.ps,height=9.5cm,angle=0}
\epsfig{figure=prz7b.ps,height=9.5cm,angle=0}
\end{center}
\caption{{\it Lines corresponding to fixed values of $\sigma
(e^+ e^- \to e^+ e^- S)$ for $\sqrt{s}=200 \gev$, in the $(m_S,
\sqrt{F})$ plane and for the parameter choices of table~1.}}
\label{sigepemS}
\end{figure}

We conclude this section with a comment similar to the one given 
at the end of sect.~2. The local effective interaction of 
eq.~(\ref{lsgsgff}) can also lead to the pair-production of 
a CP--even and a CP--odd sgoldstino, with cross-section
\be
{d \sigma \over d \cos \theta} (e^+ e^- \to S \, P ) = 
{(\tilde{m}^4_e + \tilde{m}^4_{e^c}) \over 512 \pi s^2 F^4}
\left[ (s-m_S^2-m_P^2)^2-4 m_S^2 m_P^2 \right]^{3/2} \sin^2 
\theta \, ,
\ee
where $\theta$ is the scattering angle in the center-of-mass frame.
For plausible values of the parameters, we expect this cross-section
to be suppressed by the large numerical factor and the higher power 
of the supersymmetry-breaking scale at the denominator. Otherwise, 
the corresponding signal could be seen as an anomaly in the four-jet 
sample.

\vspace{1cm}
\bc
{\bf 4. Discussion of sgoldstino signals and conclusions}
\ec

The results of sects.~2 and 3 indicate that sgoldstino production
and decay may lead to observable signals at $e^+e^-$ colliders, in
particular at LEP. In the case of $S$, the most interesting signals 
correspond to $e^+ e^- \to \gamma S$, $ZS$ or $e^+ e^- S$, followed 
by the decay $S \to gg$. Similar considerations apply to $P$. All 
these signals would deserve a dedicated experimental analysis, 
including the comparison with the SM backgrounds. In the absence of a
positive evidence, these analyses could be converted into stringent
combined bounds on the gravitino and sgoldstino masses. While 
waiting for such an experimental study, we can only give a tentative
picture of the LEP discovery potential for sgoldstinos, summarized
in fig.~8. In drawing this picture, we just assumed some 
representative values for the relevant branching ratios and 
cross-sections. The choice of $BR(Z \to S \gamma)=10^{-5}$ 
can be partially justified on the basis of some existing 
OPAL and L3 studies \cite{lep1}, applicable in the mass 
region $20 \gev < m_S < 80 \gev$. As for the other processes,
we are not aware of experimental studies whose results could
be directly applied. In the case of $\gamma S$ production,
one should generalize the LEPI analyses of \cite{lep1}
along the lines of \cite{lep2hg}. In the case of $ZS$ 
production, one could exploit some similarities with SM Higgs 
searches (see, e.g., \cite{lep2hz} and references therein), 
keeping in mind two very important differences: {\it i)} the 
production cross-section depends not only on $m_S$, but also on 
the mass parameters of table~1 and on $\sqrt{F}$, so the 
one-to-one correspondence between mass and production 
cross-section, valid for the SM Higgs, is in general lost; 
{\it ii)} massive sgoldstinos decay into gluon jets, not into $b$-jets, 
thus one cannot exploit all the machinery of $b$-tagging techniques.
Finally, in the case of $e^+ e^- \to e^+ e^- S$, we expect both
leptons to go undetected along the beam pipe in most cases,
in accordance with the validity of the approximation used in the
computation. Therefore, this signal will presumably suffer from
a larger SM background than the previous ones, where the sgoldstino
is produced in association with a photon or a $Z$. Also, we expect 
the LEP sensitivity to vary strongly with the sgoldstino masses, 
the most difficult region being the one with $m_S \sim m_Z$. 
Because of these problems, we limited ourselves to plotting contours 
of $\sigma = 10^{-1} \pb$ for the processes with a photon or a $Z$
in the final state, and of $\sigma = 1 \pb$ for the signal corresponding
to $\gamma \gamma$--fusion, leaving a detailed analysis to our 
experimental colleagues.
\begin{figure}[htbp]
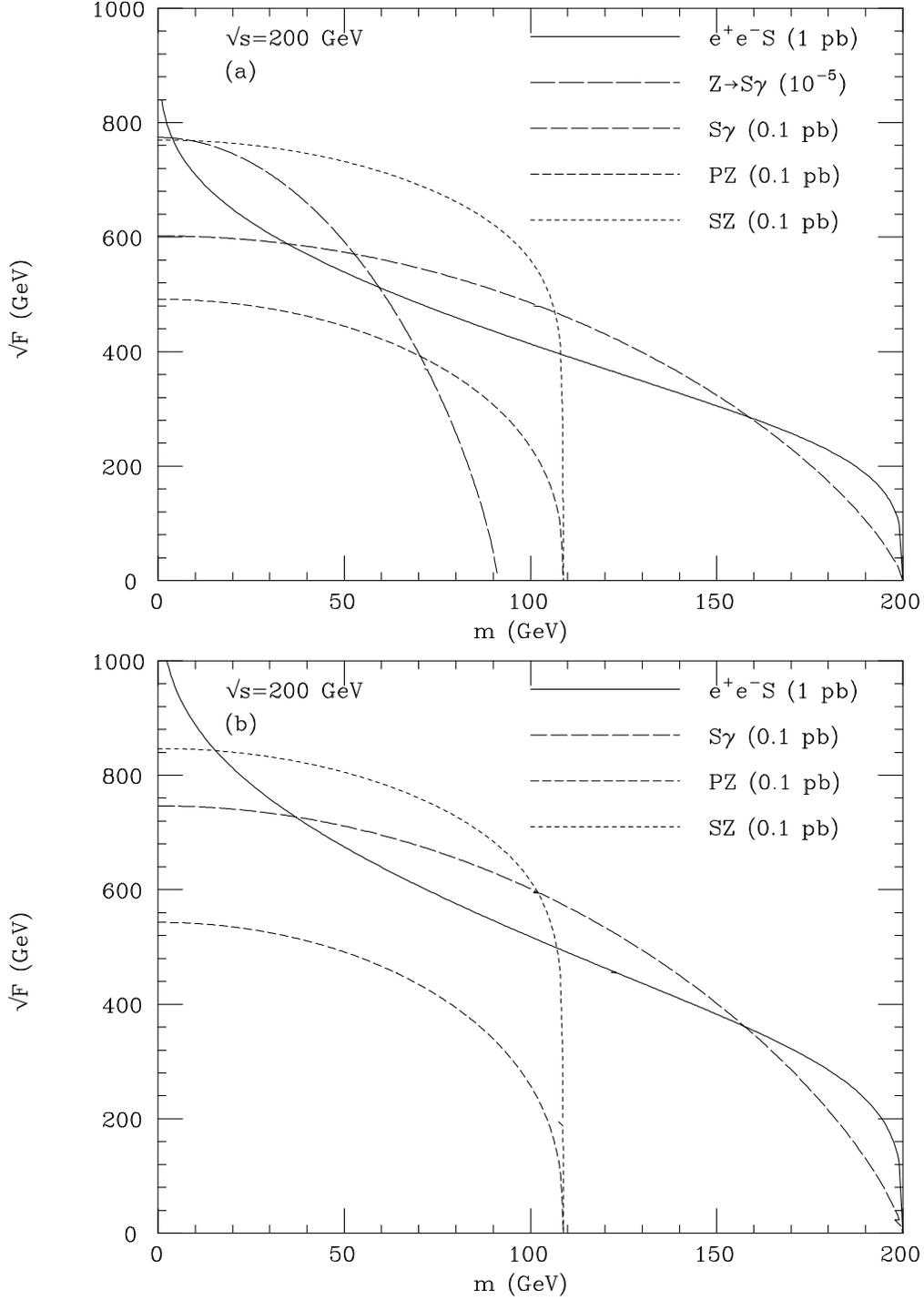

\bc
\epsfig{figure=prz8a.ps,height=9.5cm,angle=0}
\epsfig{figure=prz8b.ps,height=9.5cm,angle=0}
\ec
\caption{{\it A tentative pictorial summary of the LEP 
discovery potential in the $(m_{S(P)},\sqrt{F})$ plane, 
for the parameter choices of table~1.}}
\label{lepsum}
\end{figure}

We can see from fig.~8 that sgoldstino searches at LEP are 
likely to explore virgin land in the parameter space of models
with a superlight gravitino. For example, the present limit 
on $\sqrt{F}$, for heavy sgoldstinos and MSSM sparticles,
is only slightly above $200 \gev$ \cite{bfzph}. The associated 
production of MSSM sparticles and gravitinos is only relevant for 
masses of the MSSM sparticles smaller than the values assumed here. 
Indirect constraints from the muon anomalous magnetic moment 
\cite{bpz}, electroweak precision data \cite{bfprz} and anomalous 
four--fermion interactions \cite{bfzff} just give complementary 
and comparable bounds. Unitarity bounds just require the 
supersymmetry-breaking masses to be smaller than ${\cal O} (2$-$3) 
\times \sqrt{F}$, a condition which is comfortably fulfilled along 
most of our sensitivity contours: the only problematic regions are 
those very close ($\simlt 5 \gev$) to the boundary of the phase space 
for the process under consideration, which should therefore be excluded 
from the analyses and left for investigations at higher energies.
Another fact emerging from fig.~8 is the complementarity of the
different signals: their relative importance will depend not only
on the experimental sensitivity, but also on the specific values
of the gaugino and higgsino masses.

Finally, even if the present work is focused on sgoldstino signals 
at $e^+ e^-$ colliders, we would like to add a few comments on 
the possibility of producing sgoldstinos at hadron colliders.
At leading order in the strong interactions, the dominant 
production mechanism for massive sgoldstinos should be 
gluon-gluon fusion, since direct couplings to quark-antiquark
pairs are suppressed by the corresponding quark masses. For
sufficiently heavy sgoldstinos, the resulting signal would be 
a peak in the di-jet invariant mass distribution. Such a signal
is not present in the limit of negligible sgoldstino masses, and
was therefore neglected in previous studies \cite{dnw,dn}.
Another possibility would be to look for sgoldstino production 
in association with an electroweak gauge boson $(\gamma,Z,W)$
or a jet. In the first case, the relevant partonic processes
are $q \ov{q} \to \gamma S$, $q \ov{q} \to Z S$, $q \ov{q'}
\to W S$, whose cross-sections are the obvious generalizations 
of those given here for $e^+ e^- \to \gamma S$ and $e^+ e^- \to 
Z S$. In the second case, there are several diagrams that may 
contribute at the same order in the strong interactions, thus 
the calculation of the cross-section is considerably more 
complicated than the existing ones \cite{dn}, performed in 
the special case of negligible sgoldstino masses. We expect 
these processes to give constraints comparable with,
and complementary to, the processes considered in the present
paper. However, the theoretical complications due to the
hadronic initial state and the presence of large SM 
backgrounds require a careful analysis: we are planning to 
come back to all this in a future publication.

\vspace*{0.5cm}
{\bf Acknowledgements. }
We would like to thank S.~Ambrosanio, A.~Brignole, P.~Checchia, 
F.~Feruglio, J.-F.~Grivaz and B.~Mele for discussions and 
suggestions. One of us (F.Z.) also thanks ITP, Santa Barbara,
where part of this work was done, for the kind hospitality.
This research was supported in part by the National Science
Foundation under Grant No. PHY94-07194.
\vspace*{0.5cm}
{\bf Note added. } After the submission of the present paper, 
we were informed by P.~Checchia and G.~Wilson that, especially 
when the sgoldstino mass is close to $m_Z$, the study of the 
two-photon decay channel may lead to a sensitivity comparable 
with the two-gluon decay channel \cite{wilson,checchia}.

\end{document}